\documentclass[aps,twocolumn,groupedaddress,floatfix,showpacs]{revtex4}

\usepackage{epsfig}
\usepackage{amssymb}
\usepackage{amsmath}

\begin{document}

\title{Dissipation in a rotating frame: master equation,\\
 effective temperature and Lamb-shift}

\author{Alvise Verso and Joachim Ankerhold}

\affiliation{Institut f\"ur Theoretische Physik, Universit\"at Ulm, Albert-Einstein-Allee 11,
89069 Ulm, Germany}

\date{\today}

\begin{abstract}
Motivated by recent realizations of microwave-driven nonlinear resonators in superconducting circuits, the impact of environmental degrees of freedom is analyzed as seen from a rotating frame. A system plus reservoir model is applied  to consistently derive in the weak coupling limit the master equation for the reduced density in the moving frame and near the first bifurcation threshold. The concept of an effective temperature is introduced to analyze to what extent a detailed balance relation exists. Explicit expressions are also found for the Lamb-shift. Results for ohmic baths are in agreement with experimental findings, while for structured environments population inversion is predicted that may qualitatively explain recent observations.
\end{abstract}

\maketitle

The prospect to tailor quantum devices on ever growing scales has stimulated major experimental research in the last years. In solid state physics various types of nonlinear resonators have been fabricated such as mechanical beams \cite{cleland_03} and superconducting tunnel junctions embedded in cavities \cite{metcalfe_07}, partly to study fundamental physical phenomena, partly to develop highly sensitive amplifiers for detection schemes towards the quantum limit. These systems can be tuned over broad ranges of parameter space and particularly between the domains of classical and quantum behavior since they interact inevitably with surrounding degrees of freedom. Theory is challenged as now dynamical processes like bifurcation, parametric amplification, and period doubling appear on a level where $\hbar$ tends to play a decisive role.

A particular example of this type are microwave-driven Josephson junctions (JJ), recently realized in form of the Josephson Bifurcation Amplifier \cite{siddiqi_04,siddiqi_05} and the Cavity Bifurcation Amplifier \cite{metcalfe_07}. They are based on the fact that near the first bifurcation threshold the switching between the two driving-induced states is extremely sensitive to parameters of the JJ. The latter ones may be influenced by the coupling to a quantum system of interest, e.g.\ a Cooper pair island acting as artificial atom \cite{vion_02}, meaning that the state of the quantum system can be retrieved from the measurement of the switching rate of the JJ. It has been shown that this strategy allows for a single shot read-out close to the quantum non-demolition (QND) limit \cite{boulant_07}.
Theoretically, a general approach is provided by the Floquet representation \cite{peano_05}, but in the above situation
a more powerful procedure for analytical investigations is to describe the driven dynamics in a frame rotating with a frequency equal to the response frequency of the system as already analyzed in the classical regime by Dykman and co-workers \cite{dykman_80,dykman_98}. The extension to the quantum regime has been given in \cite{marthaler_06,dykman_07,serban_07}. In essence, one arrives at a time-independent description with a non-standard Hamiltonian though. A complete understanding, however, must include also the bath degrees of freedom residing in the lab frame.

The common formulation for dissipative systems in the weak coupling regime is provided by so-called master equations \cite{gardiner_01,breuer_02}. It is well established that the impact of the bath is twofold, namely, on the one hand  to induce transitions between eigenstates of the bare system so that they are asymptotically populated according to the temperature of the bath, and on the other hand to shift oscillation frequencies of the off-diagonal elements of the density matrix, an effect known as Lamb-shift. For driven systems the concept of an effective temperature has been introduced in \cite{marthaler_06}, but it was shown only in certain limits \cite{dykman_07,thorwart_09} that this temperature significantly differs from that in the laboratory. In these and related works \cite{serban_07,nakano_09} the focus has been on the switching between driving induced states, while an analysis of the structure of the  dissipative dynamics in the rotating frame has seen less attention. Here we fill this gap in consistently deriving a corresponding master equation and,  importantly for ongoing experimental activities, in providing an analysis to what extent a unique effective temperature can actually be introduced. This issue is intimately related to the existence of a detailed balance relation in the rotating frame.

\section{Standard formulation}

In the microscopic model for quantum dissipation a quantum mechanical system is coupled to a thermal bath so that the total Hamiltonian takes the form $ H=H_S+H_B+H_{I}$ with
 $H_S$ being the Hamiltonian of the bare system, $H_B$ the Hamiltonian of the heat bath, and $H_{I}$ the interaction, i.e.,
\begin{eqnarray}
H_B&=&\sum_{n=1}^N\frac{p_n^2}{2m_n}+\frac{m_n}{2}\omega_n^2 x_n^2\nonumber\\
H_{I}&=&-q\sum_{n=1}^N c_nx_n+q^2\sum_{n=1}^N\frac{c^2_n}{2 m_n \omega_n^2}\, .
\label{intstandard}
\end{eqnarray}
The time evolution of the density matrix of the full compound $W(t)$ obeys the Liouville-von Neumann equation $i\hbar d{W}(t)/dt=[H,W(t)]$
with an initial state $W(0)$.
The relevant operator is the reduced density $\rho(t)={\rm tr}_B\{W(t)\}$ for which a simple equation of motion does in general not exist. In case of weak friction and sufficiently fast bath modes, however, progress is made within a Born-Markov approximation. One then obtains
the usual master equation which we cast in the form
\begin{equation}\label{master}
i\hbar \frac{d\rho(t)}{dt}=[H_S,\rho(t)]-i {\cal L}_{qq}[\rho]
\end{equation}
 where
\begin{eqnarray}\label{eq1}
{\cal L}_{qq}[\rho]&=&\int_0^{\infty}ds\, K^{\prime}(s)\left[q,\left[q(-s),\rho(t)\right]\right]\nonumber\\
&&+\int_0^{\infty}ds\, i K^{\prime\prime}(s)\left[q,\lbrace q(-s),\rho(t)\rbrace\right]\,
\end{eqnarray}
contains the position operator  $q(s)$  in the interaction representation and $\{,\}$ denotes the anti-commutator.
The effective impact of the bath appears as a force-force correlator $K(t)=1/\hbar\langle \xi(t)\xi(0)\rangle_\beta=K'+iK''$ with $\xi=\sum c_n x_n$,
\begin{equation}\label{ff}
K(t)= \int_{0}^{\infty}\!\!\!\frac{d\omega}{\pi} {I}(\omega)\left[\coth\left(\frac{\omega\hbar\beta}{2}\right)\cos(\omega t)-i\sin(\omega t)\right]
\end{equation}
  including the spectral density $I(\omega)=\frac{\pi}{2}\sum_n\frac{c^2_n}{ m_n \omega_n}\delta(\omega-\omega_n)$ of the bath modes.
Processes mediated by these modes are revealed most clearly in the representation of the master equation using the eigenstates of the bare system $H_S|n\rangle=E_n|n\rangle$. An additional rotating wave (or secular) approximation (RWA), where off-resonant (fast oscillating) terms in (\ref{master}) are neglected, gives
\begin{equation}
\label{masterqq}
\frac{d\rho_{mn}}{dt}=-i\tilde{\omega}_{mn}\rho_{mn}+\delta_{mn}\sum_{k\neq n} \gamma_{nk}\rho_{kk}-\Gamma_{mn}\rho_{mn}\, .
\end{equation}
First, the reservoir induces transitions between eigenstates  with rates $\gamma_{nm}=|\langle n|q|m\rangle|^2\, D(\omega_{nm})$ at frequencies $\omega_{nm}=(E_n-E_m)/\hbar$ where
\begin{equation}
\label{eq:dfunction}
D(\omega)=\hbar\int_{-\infty}^{\infty}ds\, K(s){\rm e}^{i\omega s}=\hbar I(\omega)\, n_\beta(\omega)\,
\end{equation}
with  $n_{\beta}(\omega)=1/[\exp(\beta \hbar\omega)-1]$. The decay rates for the off-diagonal elements of the density matrix
\begin{equation}
\Gamma_{mn}=\frac{\Gamma_{mm}+\Gamma_{nn}}{2}+\frac{D(0)}{2\hbar^2}(\langle m|q|m\rangle-\langle n|q|n\rangle)^2
\end{equation}
contain the collective rates  $\Gamma_{mm}=\sum_{k\neq m} \gamma_{km}$.
Second, oscillation frequencies are renormalized $\tilde{\omega}_{nm}=\omega_{nm}+\Delta\omega_{nm}$ by the so-called Lamb-shift
\begin{equation}
\label{lambshift}
\Delta\omega_{mn}=\sum_{k}\left[-|\langle m|q|k\rangle|^2 \frac{D^L(\omega_{km})}{2\hbar}+(m\to n)\right]\, .
\end{equation}
The bath appears in form of
\begin{eqnarray}
\label{lambdl}
D^L(\omega)&=&2\int_0^{\infty}dt\Big[K'(t)\sin(\omega t)-K''(t)\cos(\omega t)\Big]\\
&=&2K'_{\rm sin}(\omega)-M[\omega\,\eta_{\rm sin}(\omega)-\eta(0)]
\end{eqnarray}
determined by
\begin{equation}
K'_{\rm sin}(\omega)=\int_0^\infty dt K'(t) \sin(\omega t)\, .
\label{ksinus}
\end{equation}
Moreover, $M$ denotes the mass of the system particle and $\eta_{\rm sin}$ is the $\sin$-transform of the classical friction kernel $\eta(t)=\int d\omega I(\omega) \cos(\omega t)/M\pi \omega$.

Since the environment rests in thermal equilibrium its probabilities $D(\omega)$ to emit and $D(-\omega)$ to absorb photons are related by detailed balance, i.e.,
\begin{eqnarray}\label{temp_ef}
e^{-\beta\hbar\omega}\equiv\frac{D(-\omega)}{D(\omega)}\, , \ \omega\geq 0\, .
\end{eqnarray}
This latter condition can also be used to define the temperature $\beta=1/k_{\rm B} T$ imposed asymptotically (for long times) by the bath onto the thermal state of the system: the stationary populations $P_n=\rho_{nn}$ of the system obey  $P_n/P_k=D(-\omega_{nk})/D(\omega_{nk})$, thus leading to the known Boltzmann distribution.

\section{Master equation in a rotating frame}

We now consider a system with a standard Hamiltonian $H_S=p^2/2M + V(q)$ where in
 \[
 V(q)=\frac{M\omega_0^2}{2}\, q^2+V_1(q)
 \]
 a harmonic term is splitted off from anharmonic contributions. This system is subject to external periodic driving with frequency $\omega_D$. Then, asymptotically the reduced density does not reach thermal equilibrium, but rather approaches a time dependent state with periodicity $T_F\equiv 2\pi/\omega_F$. For instance, the standard dipole coupling between system and driving force $\propto q \cos(\omega_D t)$\cite{peano_05} leads to $\omega_F=\omega_D$, while a parametric driving of the form $\propto \cos(\omega_D t) q^2$ \cite{marthaler_06} gives rise to period doubling $\omega_F=\omega_D/2$. As already noted above, particularly the former situation is relevant for mesoscopic detectors built with microwave-driven JJ. The domain where the detector is most sensitive, is located around the first bifurcation threshold, where the anharmonicity $V_1$ is still weak and $(\omega_0-\omega_D)/\omega_0\ll 1$.
When amplitude and/or frequency of the external drive are tuned into this regime, after a transient period of time the density takes the form $\rho(t)\sim {\bar{\rho}}(t)\cos(\omega_F t)$
with $\bar{\rho}$ changing only slowly in time, i.e., on time scales $t\gg 1/\omega_F$. In principle, to derive the master equation for $\bar{\rho}$, one may follow the strategy outlined above, however, only after switching to a moving frame. We mention in passing that this mapping  is closely related to a representation within Floquet theory \cite{peano_05} combined with a projection onto the subspace of the leading harmonics $\sim \exp(\pm i \omega_F t)$.

For this purpose, we introduce a unitary operator of the composite system
\begin{equation}
\label{eq:unitary}
U(t)\equiv U_S(t)U_B(t)=e^{-i\hat{a}^{\dag}\hat{a}\omega_F t-i \sum_n^N\hat{b}_n^{\dag}\hat{b}_n\omega_F t}\,,
\end{equation}
where $\hat{a}$ and $\hat{b}_n$ are annihilation operators for {\em harmonic} oscillators in the system and in the bath, respectively. The total Hamiltonian in the rotating frame
\[
\tilde{H}=U^{\dag}\left[H-i\hbar\frac{\partial}{\partial t}\right] U=\tilde{H}_S+\tilde{H}_B+\tilde{H}_{I}\,
\]
follows upon discarding fast oscillating terms  $\exp(\pm i k\omega_F t)\, , |k|\geq 1$ as a {\em time-independent} Hamiltonian of the form $\tilde{H}_S(Q,P)=U_S^{\dag}{\cal H}_S\, U_S$ and
\begin{eqnarray}
\tilde{H}_B&=&\sum_{n=1}^N\frac{{p}_n^2}{2\tilde{m}_n}+\frac{\tilde{m}_n}{2}\tilde{\omega}_n^2 {x}_n^2\nonumber\\
\tilde{H}_{I}&=&-\sum_{n=1}^N \tilde{c}_n\left({x}_n {Q}+\frac{{p}_n}{\tilde{\omega}_n\tilde{m}_n}\frac{{P}}{\omega_F M}\right)\nonumber\\
&+&\left({Q}^2+\frac{{P}^2}{\omega_F^2 M^2}\right)\sum_{n=1}^N\frac{{c}^2_n}{4 m_n \omega_n^2}\, .
\label{introt}
\end{eqnarray}
The operator ${\cal H}_S=H_S-i\hbar\partial_t$ coincides with the extended Hamiltonian in the Floquet description so that the spectrum of  $\tilde{H}_S$ reproduces the corresponding Floquet quasi-energies. Its eigenstates $|\tilde{\psi}_n\rangle$ are related to the Floquet states $|{\psi}_n\rangle$ via $|\tilde{\psi}_n\rangle = P_F U_S |{\psi}_n\rangle$ with $P_F$ being the projector onto the $(k\cdot\omega_F)$-subspace with $k=\pm 1$. We emphasize that the original Hamiltonian (\ref{intstandard}) cannot be regained from (\ref{introt})  in the limit $\omega_F\to 0$ since then fast oscillating terms neglected in the derivation of (\ref{introt}) contribute.

For the derivation of the master equation the specific form of $\tilde{H}_S$ is not relevant. Important are the new bath parameters
\[
\tilde{m}_n=\frac{m_n}{1-\omega_F/\omega_n}\, ,\ \tilde{\omega}_n=\omega_n-\omega_F\, , \ \tilde{c}_n=\frac{c_n}{2}\, .
\]
 Further, in the rotating frame the system-bath coupling is not just a position-position interaction but includes also the momenta, a point which we will discuss in more detail below.
Starting now with $i\hbar d\tilde{W}/dt=[\tilde{H},\tilde{W}]$ and following the standard procedure, one obtains the master equation in the rotating frame
\begin{equation}
i\hbar\frac{d\bar{\rho}}{dt}=[\tilde{H}_S,\bar{\rho}]
+\left({{\cal L}}_{QQ}+{{\cal L}}_{QP}+{{\cal L}}_{PQ}+{{\cal L}}_{PP}\right)[\bar{\rho}]\, .
\label{eq:master_rot}
\end{equation}
Here operators ${{\cal L}}_{QQ}$ and ${{\cal L}}_{PP}$ are defined according to (\ref{eq1}) with $q$ replaced by ${Q},{P}/(\omega_F M)$ with the force-force correlator defined by
\begin{eqnarray}
 \tilde{K}_{xy}&=&\tilde{K}'_{xy}+i\tilde{K}''_{xy}\nonumber\\
&=&\frac{1}{\hbar}\langle F_x(t)\,F_y(0)\rangle_\beta\hspace{.8cm} \text{with}\,\, x,y=Q,P
\end{eqnarray}
where
\begin{eqnarray}
F_Q=\sum \tilde{c}_n x_n\,,\hspace{1cm}F_P=\sum \tilde{c}_n \frac{p_n}{\tilde{\omega}_n\tilde{m}_n}\,.
\end{eqnarray}
Whereas the mixed operators are
\begin{eqnarray}\label{qp}
{\cal L}_{QP}[\bar{\rho}]&=&\frac{1}{\omega_F M}\int_0^{\infty}ds\, \tilde{K}^{\prime}_{QP}(s)\left[Q,\lbrace P(-s),\rho(t)\rbrace\right]\\
&&+\frac{1}{\omega_F M}\int_0^{\infty}ds\, i \tilde{K}^{\prime\prime}_{QP}(s)\left[Q,\left[ P(-s),\rho(t)\right]\right]\,\nonumber
\end{eqnarray}
and ${\cal L}_{PQ}$ is defined according to (\ref{qp}) with $Q\leftrightarrow P$.
The force-force correlators can be expressed as
\begin{eqnarray}
&&\tilde{K}_{QQ}(t)=\tilde{K}_{PP}(t)=\\ &&\int_{-\omega_F}^{\infty}\frac{d\omega}{\pi}\tilde{I}(\omega)\Big\{\coth\left[\frac{(\omega+\omega_F)\hbar\beta}{2}\right]\cos(\omega t)-i\sin(\omega t)\Big\}\,\nonumber
\label{eq:k_rot}
\end{eqnarray}
and
\begin{eqnarray}
&&\tilde{K}_{QP}(t)=-\tilde{K}_{PQ}(t)=\\ &&\int_{-\omega_F}^{\infty}\frac{d\omega}{\pi}\tilde{I}(\omega)\Big\{\coth\left[\frac{(\omega+\omega_F)\hbar\beta}{2}\right]\sin(\omega t)
+i\cos(\omega t)\Big\}\,\nonumber
\label{eq:k_rot2}
\end{eqnarray}
with the spectral density in the rotating frame
\[
\tilde{I}(\omega)=I(\omega+\omega_F)/4\, .
\]
Note that the unitary transformation (\ref{eq:unitary}) does not affect the equilibrium density of the bath since $[U_B,H_B]=0$, but just the dynamics of the correlator. Accordingly, equilibrium properties in (\ref{eq:k_rot}) appear through a frequency shift $\omega\to \omega-\omega_F$ meaning that in the rotating frame the bath carries modes with "negative" frequencies. The above analysis shows that in the rotating frame a typical friction strength is given by $\tilde{\eta}={I}(\omega_F)/4M\omega_F$ which is smaller than that in the lab frame $\eta=\lim_{\omega\to 0}I(\omega)/M\omega$. The frequency shift in the bath correlation function $\tilde{K}(t)$ effectively produces a decay in time which qualitatively is similar to that of $K(t)$. The condition for the validity  of the master equation (\ref{eq:master_rot}) can thus be estimated to read $\tilde{\eta}\hbar\beta\ll 1$ provided a typical bath cut-off frequency $\omega_c$ sufficiently exceeds $\eta>\tilde{\eta}$.

It is instructing to express the above master equation in terms of annihilation and creation operators. This is most conveniently done in the interaction picture representation where the reduced density is given by $\bar{\rho}_I(t)=e^{i\tilde{H}_S t/\hbar}\bar{\rho}(t) e^{-i\tilde{H}_S t/\hbar}$. This way, (\ref{eq:master_rot}) translates into
 \begin{equation}
i\hbar\frac{d\bar{\rho}_I}{dt}=
\left({{\cal L}}^I_{QQ}+{{\cal L}}^I_{QP}+{{\cal L}}^I_{PQ}+{\cal L}^I_{PP}\right)[\bar{\rho}_I]\, .
\label{eq:master_rot2}
\end{equation}
We now introduce  operators
\begin{eqnarray}
a(\omega)&=&\sum_{E^{\prime}-E=\hbar\omega}|E\rangle\langle E|a|E^{\prime}\rangle\langle E^{\prime}|\\
a^{\dagger}(\omega)&=&\sum_{E-E^{\prime}=\hbar\omega}|E\rangle\langle E|a^{\dagger}|E^{\prime}\rangle\langle E^{\prime}|\, ,
\end{eqnarray}
where the sum runs over all energy eigenstates $|E\rangle$, $|E'\rangle$ of $\tilde{H}_S$ with fixed energy difference $\hbar\omega$.
$a$ and $a^{\dagger}$ are annihilation and creation operators of a harmonic oscillator with frequency $\Omega$, i.e.,
\begin{eqnarray}
\label{creation}
a&=&\sqrt{\frac{M\Omega}{2\hbar}}\left(Q+\frac{i P}{M\Omega}\right)\nonumber\\ a^{\dagger}&=&\sqrt{\frac{M\Omega}{2\hbar}}\left(Q-\frac{i P}{M\Omega}\right)\, .
\end{eqnarray}
The above frequency dependent operators satisfy $[\tilde{H}_S, a^\dagger(\omega)]=\hbar\omega a^\dagger(\omega)$ and  $[\tilde{H}_S, a(\omega)]=-\hbar\omega a(\omega)$ and are thus creation and annihiliation operators at frequency $\omega$.
Note that these relations hold independent of the frequency $\Omega$ in the definition (\ref{creation}). For a purely harmonic system with frequency $\tilde{\omega}_0$, however, the  $a^\dagger(\omega), a(\omega)$ reduce to  $a, a^\dagger$ only if $\Omega=\tilde{\omega}_0$. It is thus convenient to chose $\Omega$ as the frequency of small oscillations around one of the stable extrema of $\tilde{H}_S$, for instance that one where the dynamics starts initially.

The sum of dissipative operators in (\ref{eq:master_rot2}) can now be expressed as
\begin{equation}
{\cal L}_{QQ}^I+{\cal L}_{QP}^I+{\cal L}_{PQ}^I+{\cal L}_{PP}^I=\frac{\hbar}{2M\Omega}\sum_{\omega, \omega'} L_{\omega, \omega'}
\end{equation}
with
\begin{align}\nonumber
&L_{\omega, \omega'}[\bar{\rho}_I]={\rm e}^{-it(\omega-\omega^{\prime})}\times \nonumber\\
&\Big\{\hat{K}_{QQ}'(\omega')[a(\omega),[\lambda^-\, a^{\dagger}(\omega^{\prime})+\lambda^+ a(-\omega^{\prime}),\bar{\rho}_I(t)]]\nonumber\\
&+i\mu\hat{K}_{QP}'(\omega')[a(\omega),\lbrace\, a^{\dagger}(\omega^{\prime}),\bar{\rho}_I(t)\rbrace]\nonumber\\
&+i\hat{K}_{QQ}''(\omega')[a(\omega),\lbrace\lambda^-\, a^{\dagger}(\omega^{\prime})+\lambda^+\, a(-\omega^{\prime}),\bar{\rho}_I(t)\rbrace]\nonumber\\
&-\mu\, \hat{K}_{QP}''(\omega')[a(\omega),[\, a^{\dagger}(\omega^{\prime}),\bar{\rho}_I(t)]]\Big\}+h.c.
\label{sumop}
\end{align}
Here, the Laplace transforms of the damping kernel (\ref{eq:k_rot}) follow from
\begin{eqnarray}
\hat{K}_{xy}'(\omega)=\int_0^{\infty}dt\, \tilde{K}_{xy}^{\prime}(t){\rm e}^{-it\omega}\nonumber\\
\hat{K}_{xy}''(\omega)=\int_0^{\infty}dt\, \tilde{K}_{xy}''(t){\rm e}^{-it\omega}
\end{eqnarray}
and the impact of the momentum dependent coupling terms between system and bath are taken into account by coefficients
\begin{equation}
\mu=\frac{2\Omega}{\omega_F}\ , \ \lambda^{\pm}=1\pm\frac{\Omega^2}{\omega_F^2}\, .
\end{equation}
As for the standard case, a further simplification arises  within the RWA, where all terms in the sum (\ref{sumop}) with $\omega\neq \omega'$ are assumed to oscillate so rapidly on the time scale for relaxation that they may be discarded.  Accordingly, (\ref{sumop}) reduces to
\begin{equation}\label{lq}
\left.{\cal L}_{QQ}^I+{\cal L}_{PQ}^I+{\cal L}_{QP}^I+{\cal L}_{PP}^I\right|_{RWA}=\frac{\hbar}{2 M \Omega}\sum_{\omega} L_{\omega,\omega}
\end{equation}
We recall that the anharmonicity of the system potential gives rise to the frequency sum. In contrast,
for a purely harmonic system with frequency $\Omega=\tilde{\omega}_0$ it collapses to  contributions for $\omega=\pm \Omega$. Moreover, $a(-|\omega|)=a^\dagger(-|\omega|)=0$ so that terms carrying in (\ref{sumop}) the coefficient $\lambda^+$ vanish. One should emphasize that this is {\em not} the case if in (\ref{creation}) a frequency $\Omega\neq \tilde{\omega}_0$  is chosen. A conventional master equation then emerges with renormalized friction functions $\hat{K}'\to  (\lambda^-\hat{K}_{QQ}'-\mu\hat{K}_{QP}'')$ and $\hat{K}''\to  (\lambda^-\hat{K}_{QQ}''+\,\mu\hat{K}_{QP}')$. In the Wigner representation the momentum coupling terms ensure that in the laboratory frame the known Fokker-Planck equation for driven systems is reproduced. To neglect them in the rotating frame leads in turn to spurious diffusion terms in the lab frame \cite{kohler_97}. 

In the anharmonic case and for elevated temperatures it may be more convenient not to work explicitly with annihilation/creation operators, but rather use a representation as in (\ref{masterqq}), where matrix elements of those system operators appear which couple to the bath. In particular, these matrix elements can then directly be evaluated for the full anharmonic problem by means of semiclassical techniques in certain ranges of parameter space \cite{marthaler_06,verso_09}.
The standard result can easily  be generalized to the rotating frame situation: in addition to matrix elements $\langle m|Q|n\rangle $ also $\langle m|P/M\omega_F|n\rangle $ must be taken into account; these matrix elements are multiplied by the bath correlation functions $\tilde{D}_{xy}$ and $\tilde{D}_{xy}^L$ derived from $\tilde{K}_{xy}$. The diagonal part of the density $\bar{\rho}$ thus obeys a Pauli-master equation with properly modified transition rates,
\begin{eqnarray}
\gamma_{nm}=\tilde{D}_{QQ}(\omega_{mn})\left(|\langle n|Q|m\rangle|^2+\frac{1}{(\omega_FM)^2}|\langle n|P|m\rangle|^2\right)\nonumber\\
+\frac{\tilde{D}_{QP}(\omega_{mn})}{\omega_F M}\left(\langle n|Q|m\rangle\langle m|P|n\rangle-\langle n|P|m\rangle\langle m|Q|n\rangle\right).
\end{eqnarray}
If typical transition frequencies in $\tilde{H}_S$ are small compared to the driving frequency $\omega_F$ the dominant process in the transition rates is given by the position matrix element. For instance, close to a harmonic minimum with frequency $\tilde{\omega}_0$ and near the bifurcation threshold  the momentum dependent terms provide contributions which are suppressed by factors on the order of $\tilde{\omega}_0/\omega_F\ll 1$.

As already mentioned above, the RWA does not always apply. For instance,
for higher lying states the energy spectrum of $\tilde{H}_S$ may have accumulation points with a dense distribution of transition frequencies. This issue becomes particular relevant for the situation described above, namely, a driven system close to its first bifurcation threshold. Then typically $\tilde{H}_S$ exhibits two stable domains in phase space which are separated by an unstable one, as e.g.\ for parametric driving where one has a double well structure with the two minima corresponding to the stable extrema and the barrier top to an unstable saddle point of the dynamics. Accordingly, for sufficiently deep wells the RWA applies for low lying states. By tuning the system close to a bifurcation point, however, extrema tend to coalesce so that the spectrum of $\tilde{H}_S$ becomes narrowly spaced and the RWA is no longer applicable.

\section{Effective temperature}\label{efftemp}

The model described by (\ref{introt}) can be considered as a system subject to relaxation as in the standard situation. Effectively for long times the bath dictates its temperature to the system such that the populations of the quasi-energy levels become stationary  and are distributed according to a balance between emission and absorption processes.
In case of a purely harmonic system the corresponding stationary populations can be calculated explicitly \cite{kohler_97}.
For the anharmonic case they follow within the RWA from the extended Pauli-master equation.
It is thus convenient to introduce the concept of an effective temperature, which is defined according to (\ref{temp_ef}) by
 $\exp(-\beta_{xy}\hbar\omega)=\tilde{D}_{xy}(-\omega)/\tilde{D}_{yx}(\omega)$, with $x,y=Q,P$.
Now, a quantum of energy $\hbar\omega\geq 0$ emitted from the system reaches the bath in the lab frame either with energy $\hbar\omega_+=\hbar(\omega_F+\omega)$ or $\hbar\omega_-=\hbar(\omega_F-\omega)$, thus determining two distinct regimes. In the range $\omega>\omega_F$ only $\omega_+>0$ so that only one channel of bath modes is accessible and one unique effective temperature follows from the standard expression with a shifted frequency
\begin{equation}
\beta^>\equiv\beta_{xy}^>=\beta\left(1+\frac{\omega_F}{\omega}\right)\, .
\end{equation}
The situation is different in the second range $\omega<\omega_F$ where both frequencies $\omega_\pm>0$ and {\em two} channels in the bath are open to give different effective temperatures for different diffusion processes, namely,
\begin{eqnarray}\label{efftemps}
\beta_{QQ}^<=\frac{1}{\hbar\omega}
\ln\left\{\frac{[n_{\beta}(\omega_+)+1]{I}(\omega_+)+n_{\beta}(\omega_-){I}(\omega_-)}{n_{\beta}(\omega_+) {I}(\omega_+)+
[n_{\beta}(\omega_-)+1]{I}(\omega_-)}\right\}\\
\beta_{QP}^<=\frac{1}{\hbar\omega}
\ln\left\{\frac{[n_{\beta}(\omega_+)+1]{I}(\omega_+)-n_{\beta}(\omega_-){I}(\omega_-)}{-n_{\beta}(\omega_+) {I}(\omega_+)+
[n_{\beta}(\omega_-)+1]{I}(\omega_-)}\right\}
\label{efftemps2}
\end{eqnarray}
and from (\ref{eq:k_rot}) and (\ref{eq:k_rot2}) also $\beta_{QQ}^<=\beta_{PP}^<$ and $\beta_{QP}^<=\beta_{PQ}^<$.
\begin{figure}
\epsfig{file=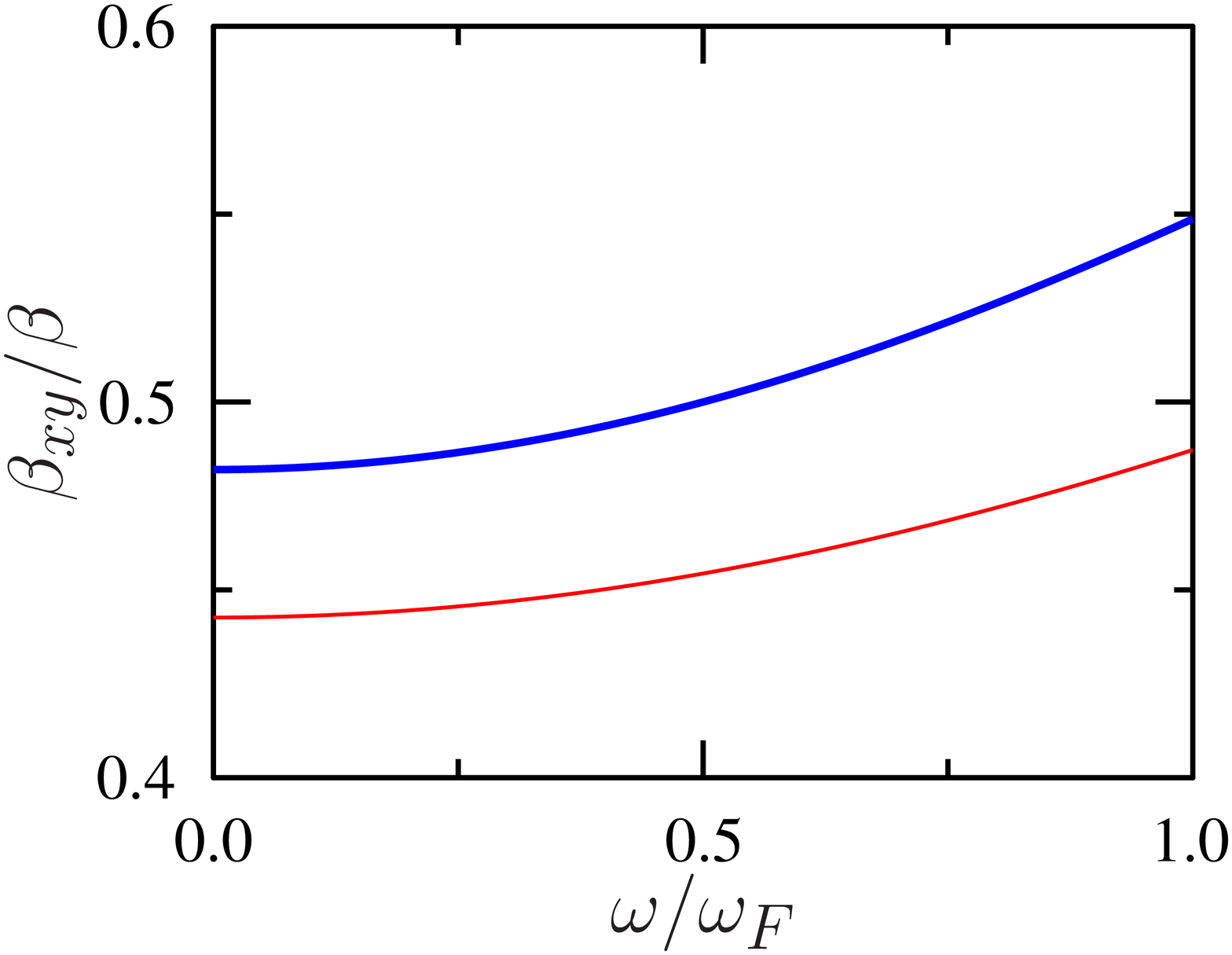, height=5.5cm}
\caption[]{\label{fig1}$\beta_{QQ}$ (thick) and $\beta_{QP}$ (thin) vs. frequency for fixed lab temperature $\beta\hbar\omega_F=4$ for an ohmic bath}
\epsfig{file=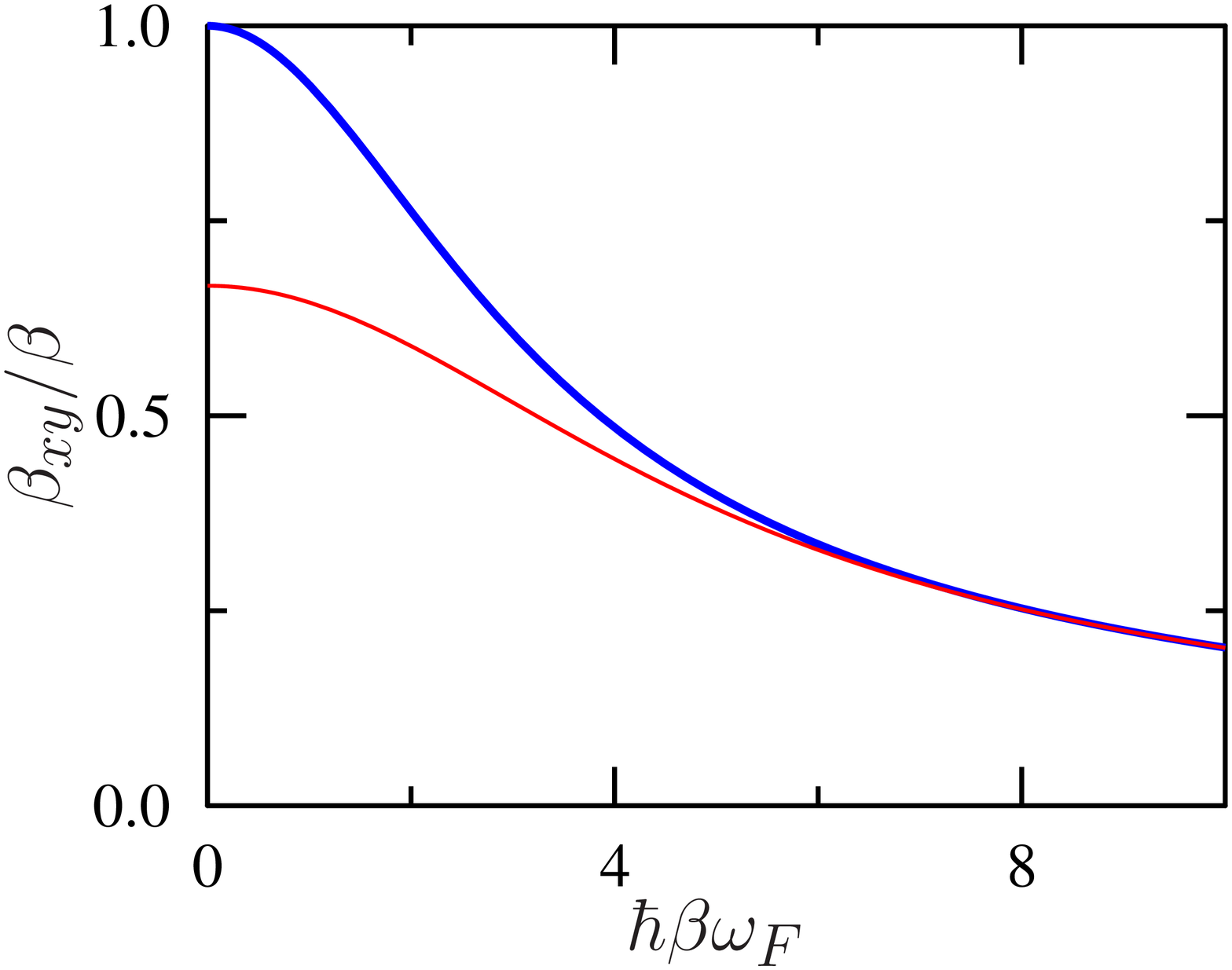, height=5.5cm}
\caption[]{\label{fig1b}$\beta_{QQ}$ (thick) and $\beta_{QP}$ (thin) vs. lab temperature at $\omega=0.2\omega_F$ for an ohmic bath.
}
\end{figure}
These results are illustrated in (fig.\ref{fig1}-\ref{fig1b}). Apparently, the different combinations  of the two channels  lead to $\beta_{QQ}^<\neq\beta_{QP}^<$ so that a unique effective temperature cannot be specified, which in turn means that a detailed balance relation does in general {\em not} exist in the rotating frame. Only if the temperature tends to zero, i.e.\ $\omega_F\hbar\beta\gg 1$, do the individual effective temperatures merge (see fig.{\ref{fig1b}}) so the detailed balance is reestablished.

Let us now discuss the situation in more detail. First, while one recovers for vanishing rotating frame frequency $T_{xy}^>=T_{xy}^<=T$, for finite $\omega_F$ a discontinuity occurs $ T_{xy}^<(\omega_F)>T_{xy}^>(\omega_F)\equiv T/2$.
Second, for $T=0$ one has $T_{xy}^>=0$,  while the  expressions (\ref{efftemps}) and (\ref{efftemps2}) predict that the common effective temperature  $T_{QQ}^<=T_{QP}^<=(\hbar\omega/k_{\rm B})/\ln[{I}(\omega_+)/{I}(\omega_-)]$ can be finite.
Third, for  small frequencies $\omega\ll\omega_F$ more transparent expressions can be derived, i.e.,
\begin{equation}\label{temp}
T_{QQ}^<=\frac{\hbar {I}(\omega_F)}{2\,k_{\rm B} I'(\omega_F)} \coth\left(\frac{\hbar\omega_F\beta}{2}\right)
\end{equation}
\begin{equation}\label{temp2}
T_{QP}^<=\frac{\hbar }{2\,k_{\rm B} } \frac{{I}(\omega_F)(e^{\hbar\omega_F\beta}-1)^2}{e^{\hbar\omega_F\beta} \hbar\omega_F\beta{I}(\omega_F)+I'(\omega_F)(e^{2\hbar\omega_F\beta}-1)}
\end{equation}
with  ${I}^{\prime}=d {I}/d\omega$. For an ohmic spectral density ${I}=M\gamma\omega$ with $I(\omega_F)/I'(\omega_F)=\omega_F$  this is in agreement with experimental observations \cite{vijayaraghavan_08}, where one always has $T_{\rm eff}^<>T$  and for $T=0$ arrives at $T_{QQ}^<=T_{QP}^<=\hbar\omega_F/2k_B$ \cite{dykman_07}. Accordingly, the rotating frame system behaves quantum mechanically only if $\hbar\omega\geq k_{\rm B} T_{xy}^<$, i.e. $\omega_F\gg \omega\geq \omega_F/2$. In contrast, in the high frequency domain the temperature is always reduced.

\begin{figure}
\epsfig{file=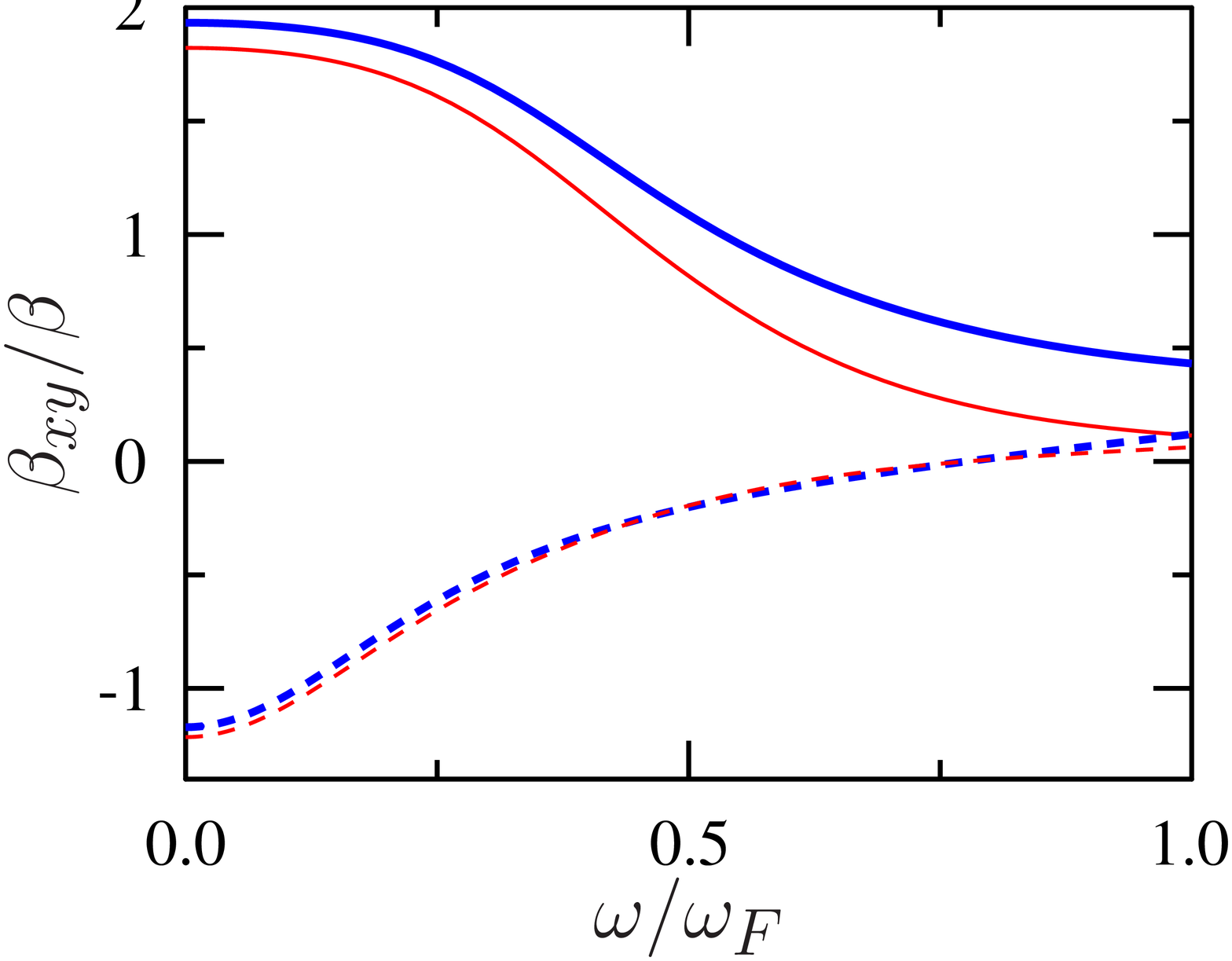, height=5cm}
\caption[]{\label{fig2}$\beta_{QQ}$ (thick) and $\beta_{QP}$ (thin) vs. frequency for fixed lab temperature $\beta\hbar\omega_F=4$ for a damped harmonic oscillator bath with $\gamma/\omega_F=0.5$ at $\omega_p=.95\omega_F$ (dashed) and $\omega_p=1.35\omega_F$ (solid).}
\epsfig{file=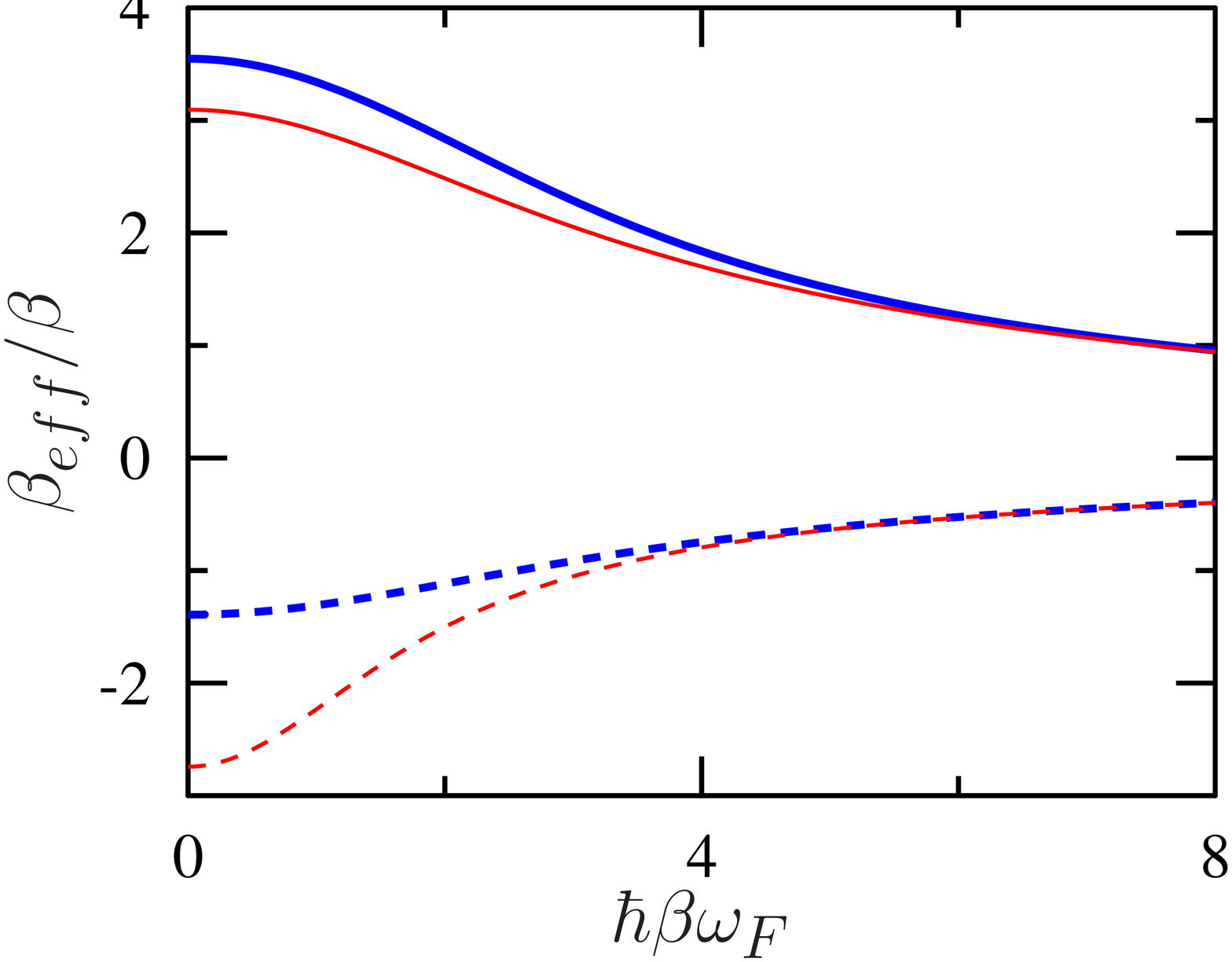, height=5cm}
\caption[]{\label{fig2b}Effective inverse temperature vs. lab temperature at $\omega=0.2\omega_F$ and for spectral bath  densities as in fig.~\ref{fig2}.}
\end{figure}

 For a structured environment the two-channel-processes in (\ref{efftemps}) can lead to counter-intuitive phenomena as seen in figs.~\ref{fig2}, \ref{fig2b}. There, we consider the coupling to a damped harmonic oscillator (frequency $\omega_p$, damping strength $\gamma$), i.e.,
 \begin{equation}
 I_p=\frac{M\gamma \omega}{(\omega^2-\omega_p^2)^2+\gamma^2\omega^2}\, ,
  \end{equation}
  for which $I'(\omega_F)<0$ if  $\omega_F>\omega_p$.
  Hence, (\ref{temp}-\ref{temp2}) predicts a {\em negative} effective temperature physically corresponding to the fact that absorption becomes more probable than emission and a population inversion is induced.
 Recent experimental observations of enhanced relaxation in a quantronium circuit coupled to a cavity bifurcation amplifier \cite{private} may be explained qualitatively by this phenomenon.
 We mention in passing that the Markov approximation employed in the derivation of (\ref{eq:master_rot}) remains valid as long as the frequency $\omega_p$ exceeds the typical frequency for the bare system dynamics $\Omega$.
\begin{figure}
\vspace*{0.4cm}
\epsfig{file=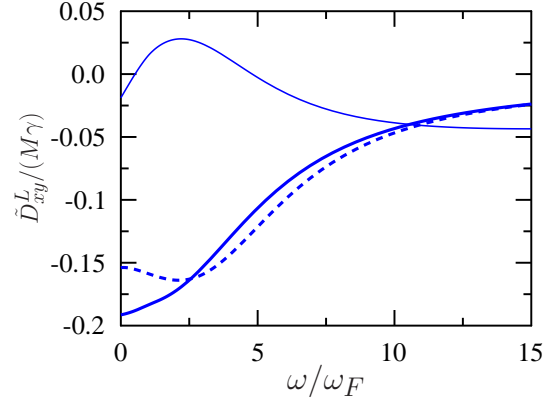, height=5.5cm}
\caption[]{\label{fig3}Bath factor for the Lamb-shift in the lab $D^L$ (dashed) and the rotating frame $\tilde{D}_{QQ}^L$ (thick) and $\tilde{D}_{QP}^L$ (thin) for $\hbar\beta\omega_F=4 $.} \vspace*{-0.25cm}
\end{figure}

\section{ Lamb-shift}
As discussed above, the reservoir also affects the frequencies of the off-diagonal elements of the density matrix. This Lamb shift can now easily be obtained from (\ref{lambshift}) as
\begin{align}
\Delta\omega_{nm}&=-\Big\{\sum_{r=0}^\infty\sum_{X,Y\in\{Q,P/M\omega_F\}}\Big[\frac{\tilde{D}_{X,Y}^L(\omega_{rm})}{2\hbar^2}\langle m|X|r\rangle\nonumber\\
&\times\langle r|Y|m\rangle\Big]-\frac{\tilde{D}_{Q,P}^L(0)}{\hbar^2}\langle n|Q|n\rangle\langle m|P/(\omega_F M)|m\rangle\nonumber\\
&-(m\to n)\Big\}
\end{align}
with the friction function 
\begin{eqnarray}
\tilde{D}_{QQ}^L&=&\tilde{D}_{PP}^L\nonumber\\
&=&2\tilde{K}'_{\rm sin}(\omega)+2\int_0^{\infty}\!\!\frac{d\Omega}{\pi} (\omega_F-\Omega)\,d_0(\Omega)
\end{eqnarray}
and
\begin{eqnarray}
&& \tilde{D}_{QP}^L(\omega)=-\tilde{D}_{PQ}^L(\omega)\nonumber\\
&&=2\int_0^{\infty}dt\Big[K_{QP}'(t)\cos(\omega t)+K_{QP}''(t)\sin(\omega t)\Big]\\
&&=2\frac{\omega_F}{\omega}\tilde{K}'_{\rm sin}(\omega)-\sum_{n=-\infty}^{\infty}\!\int_0^{\infty}\!\!\frac{d\Omega}{2\hbar\beta\pi}\,d_n(\Omega)
\nonumber\\
&&+2\int_0^{\infty}\!\!\frac{d\Omega}{\pi}\omega\, d_0(\Omega)
\end{eqnarray}
with
\begin{equation}
\tilde{K}'_{\rm sin}(\omega)=\!\sum_{n=-\infty}^{\infty}\!\int_0^{\infty}\!\!\frac{d\Omega}{4\hbar\beta\pi}
\frac{\omega}{\Omega}d_n(\Omega)
\label{ksinusrot}
\end{equation}
and
\begin{equation}
d_n(\Omega)=\frac{I(\Omega)\, \Omega^2}{\Omega^2+\nu_n^2}\frac{1}{\omega^2-(\Omega-\omega_F)^2}
\end{equation}

and the Matsubara frequencies $\nu_n=2\pi n/\hbar\beta$.
Since for a purely ohmic spectral density the above expression diverges in the low as well as in the high frequency range, we take $I(\omega)=M\gamma\omega_c^2 \omega^2/(\omega^3+\omega_c^3)$ (see fig.\ref{fig3}).
Deviations are small in the high frequency domain, but pronounced effects occur for $\omega\approx\omega_F$ such that even the sign of the respective bath factors changes.

\section{Conclusions}
To summarize we have provided a consistent derivation of a master equation needed to capture the dissipative dynamics of periodically driven nonlinear resonators near the first bifurcation. Various expressions for the corresponding master equation have been derived and discussed. It turns out that a position-position interaction between system and bath in the laboratory frame translates into additional momentum-momentum and momentum-position couplings in  the rotating frame.
In case that typical transition frequencies in the rotating frame are sufficiently small compared to the external driving frequency, these latter terms are small compared to the dominant position-position coupling.
The concept of an effective temperature has been introduced to analyze to what extent a detailed balance relation exists in the rotating frame.  In the strict sense, one recovers detailed balance only at very low temperatures.
For a structured environment phenomena such as negative effective temperatures are predicted.
Explicit expressions have also been given for the Lamb-shift, which allow to better understand recent and future experimental findings.

\acknowledgements
Fruitful discussions with P. Bertet, D. Vion, V.Peano and M. Thorwart are gratefully acknowledged. Financial support was provided by the GIF and the Landesstiftung BW.

\bibliographystyle{unsrt}
\bibliography{biblio}

\end{document}